# Transient Thermo-elasto-hydrodynamic Study of Herringbone-grooved Mechanical Face Seal during Start-up Stage

Yongfan LI[a][*], Muming HAO[b], Noël BRUNETIÈRE[c], Qiang LI[b], Jiasheng WANG[a], Baojie REN[d]

[a] College of Mechanical and Electrical Engineering, Qingdao Agricultural University, Qingdao 266109, Shandong, China

[b] College of New Energy, China University of Petroleum (East China), Qingdao 266580, Shandong, China

[c] Institut Pprime UPR 3346, CNRS - Université de Poitiers - ISAE ENSMA, Futuroscope Chasseneuil 86962, France

[d] Dongying Hiscien Sealing Technology Co. Ltd., Dongying 257000, Shandong, China

* Corresponding author. E-mail address: liyongfan@qau.edu.cn (Yongfan LI)

**Abstract:** A comprehensive numerical solution is developed for the transient thermo-elasto-hydrodynamic (TEHD) characteristics of mechanical face seals. Transient lubrication features of the fluid film, transient thermal deformation features of the seal rings, dynamic behavior, and rough faces contacting are coupled. The finite volume method is utilized for the fluid film solution, and the Duhamel's principle contributes to calculation of the time-varying solid properties. An overall flowchart for the numerical solution is established, with an approach of Parallel Dual Time Steps (PDTS approach) proposed and utilized for the explicit time solver. Both of the efficiency and accuracy of the PDTS approach are evaluated by comparing with the reference. An outer-herringbone-grooved face seal in a start-up stage is studied. The simultaneously existing physical effects of the face expansion and the seal ring movement are successfully simulated with the proposed method. Neglecting viscosity-temperature effect and convergent gap forming could underestimate the load-carrying capacity of the fluid film; smaller contacting force but larger maximum contacting pressure are found comparing with the THD and HD results; performance keeps varying at steady speed due to thermal lag effect. The proposed numerical solution could be impactful for mechanism analyzing of the undesirable running of mechanical face seals related to the transient TEHD effects.

**Nomenclature**

| | |
|---|---|
| $C$ | coefficients of the discretized Reynolds equation and energy equation |
| $c_p$ | specific heat capacity at constant pressure, [J/(kg·K)] |
| $D$ | damping of the supporting elements, [N·s/m] |
| $d$ | height of asperity after deformation, [m] |
| $F, F_o, F_c$ | resultant force, opening force and closing force acting on the flexibly mounted rotor, [N] |
| $F_{fl}, F_{ct}$ | fluid force (load-carrying capacity of the fluid film), contacting force, [N] |
| $H$ | hardness of the soft material, [Pa] |
| $h$ | nominal clearance of the sealing gap (nominal thickness of the fluid film), [m] |
| $h_g$ | depth of herringbone grooves, [m] |
| $\bar{h}_T$ | local mean clearance of the sealing gap, [m] |
| $\Delta h_u$ | clearance increment induced by face axial deformations, [m] |
| $K$ | stiffness of the supporting elements, [N/m] |
| $k$ | thermal conductivity (of the fluid film), [W/(m·K)] |
| $M$ | mass of the flexibly mounted rotor, [kg] |
| $M_{fl}, M_{ct}, M_{sum}$ | fluid, contacting and total frictional torques in the sealing gap, [N·m] |
| $N$ | groove number |
| $n$ | rotation speed, [r/min] |
| $P_{cr}$ | critical normal force of asperity, [N] |
| $P_{fl}, P_{ct}, P_{sum}$ | fluid thermal dissipation, contacting frictional power, total frictional power assumption, [W] |
| Pr | Prandtl constant |
| $p$ | fluid film pressure, [Pa] |
| $p_{ct}$ | contacting pressure of the seal faces, [Pa] |
| $p_t$ | transition pressure in the general model for gas volume fraction of the fluid film, [Pa] |
| $Q_{m,i}$ | radial mass flow rate at the inner diameter of the sealing gap, [g/min] |
| $q_\theta, q_r$ | circumferential and radial flow rates in the fluid film, [m²/s] |
| $R$ | radius of the asperity top, [m] |
| $R_{grn}, R_{grs}, R_{gre}, R_{grw}$ | groove segment length ratios, defined as ratios of the segment length in the groove and the whole length respectively for the north, south, east, west faces of each control volume in the FVM |
| $\bar{R}_{gr}$ | average groove segment length ratio |
| $R_{ldn}, R_{lds}, R_{lde}, R_{ldw}$ | $1-R_{grn}$, $1-R_{grs}$, $1-R_{gre}$, $1-R_{grw}$ |
| $r, \theta$ | polar coordinates |
| $r_{gas}$ | individual gas constant of air, 287.05 J/(kg·K) |
| $T$ | temperature (of the fluid film), [°C] |
| $t$ | time, [s] |
| $\Delta t$ | time step in the PDTS approach for the dynamic and fluid features solving, [s] |
| $u$ | axial deformation of the seal faces, [m]; radial velocity in the fluid film, [m/s] |
| $v$ | circumferential velocity in the fluid film, [m/s] |
| $w$ | normal deformation of asperity, [m] |
| $w_{cr}$ | critical normal deformation of asperity, [m] |
| $z$ | height of asperity before deformation, [m]; $z$ axis of the coordinates |
| $z_r, \dot{z}_r, \ddot{z}_r$ | axial displacement, velocity and acceleration of the reference of the flexibly mounted rotor, [m], [m/s], [m/s$^2$] |
| $z_s$ | axial displacement of the reference of the non-flexibly mounted stator, [m] |
| $\alpha$ | local gas volume fraction of the fluid film; spiral angle of herringbone grooves, [°] |
| $\eta$ | asperity amount in unit area, [1/m$^2$] |
| $\lambda$ | convective coefficient of heat transfer between the fluid film and the seal faces, [W/(m$^2\cdot$K)] |
| $\lambda_g$ | mass fraction of air in water, defined in the general model for gas volume fraction of the fluid film |
| $\mu$ | fluid viscosity, [Pa·s] |
| $\mu_{ct}$ | dry frictional coefficient of the paired materials |
| $\rho$ | fluid density, [kg/m$^3$] |
| $\rho_M$ | theoretical maximum density, defined in the general model for gas volume fraction of the fluid film, [kg/m$^3$] |
| $\sigma, \sigma_s$ | combined deviation of the asperity height distribution, [m] |
| $\phi_p$ | pressure flow factor of the average flow model |
| $\phi_{s,rs}$ | shear flow factor of the average flow model |
| $\tau$ | dummy temporal variable, [s] |
| $\omega$ | angular velocity of rotation, [rad/s] |
| **Subscript** | |
| $C$ | center node of a control volume in the FVM |
| $ct$ | representing contacting of the seal faces |
| $fl$ | representing the fluid film |
| $me1$ | representing mechanical deformation of seal faces due to face pressure |
| $me2$ | representing mechanical deformation of seal faces due to medium pressure |
| $N, S, E, W$ | north, south, east, west nodes of the center nodes in the FVM |
| $n, s, e, w$ | north, south, east, west faces of a control volume in the FVM |
| $n$ | iterative sequence of the Reynolds equation or energy equation solving; radial element amount of the faces in the FEM related to solving the solid features |
| $r$ | representing the rotor of the seal |
| $s$ | representing the stator of the seal; representing the solid features |
| *th* | representing the thermal features |
| **Superscript** | |
| *k* or (*k*) | temporal dimension sequence in the PDTS approach for the dynamic and fluid features solving |
| *l* | temporal dimension sequence in the PDTS approach for the solid features solving |

## 1. Introduction

Mechanical face seals are critical components in modern industry, preventing leakage of pressurized fluid in rotating machinery. To perform reliably under severe operating conditions, these seals depend on the formation of a micrometer-scale lubricating fluid film between the relatively moving seal faces. Extensive research has explored seal characteristics, focusing on core elements governing film thickness: including rheology and phase change, hydrodynamic and/or hydrostatic lubrication, thermo-fluid-structure coupling, contact mechanics and friction, and dynamics. Although the steady-state behavior of such seals is now well understood through extensive study, many operate under transient conditions such as repeated start-ups and shut-downs, varying fluid environments, and vibrations. Accurate simulation tools for these conditions are essential, and developing such tools is the primary goal of this paper.

Early studies on transient behavior focused on the dynamic response of mechanical seals to establish their dynamic stability limits, predominantly based on isothermal assumptions. The numerical simulation of transient response was achieved for gas-lubricated coned-face seals with flexibly mounted stators (FMS), revealing two instability modes: half-frequency whirling and face contact [1]. Responses to stator misalignment and rotor runout were also quantified [2]. The numerical formulation was presented for the spiral-grooved gas lubricated FMS seal [3], showing overly large face opening occurring in case of large spiral angle. Liu et al. [4] studied the dynamic response of wavy-tilt-dam face seal to disturbances via coupling the formula of fluid film thickness, Reynolds equation and dynamic equations. Blasiak and Zahorulko [5] did similar researching for gas face seals with several type of face modifications. Varney and Green [6] analyzed the faces impact phenomena of a coning face seal, finding aperiodic vibration and contact of faces, which shows a mechanism of the severe-contact-induced seal failure. Li et al. [7] studied the dynamic responses of a liquid lubricated spiral groove face seal to pressure variation, shaft drifting and shaft bending, emphasizing the leakage risk caused by rotor motions.

The studies mentioned above typically assumed constant seal face geometry and fluid properties, unaffected by time or operating conditions. However, thermal effects and solid deformation — critical factors governing seal behavior — can cause viscosity variations, temperature rise, and changes in film thickness. Consequently, the thermo-elasto-hydrodynamic (TEHD) lubrication performance of mechanical seals has become a major research focus. The typical double tapered hydrostatic mechanical seal in the reactor coolant pumps was studied considering deformation due to the extreme operation at high pressure, high speed and wide temperature range [8]. A 3D TEHD model of the wavy-tilt-dam mechanical seal was configured to study the mechanisms under the steady-state and quasi-start-up conditions [9][10]. The mixed TEHD model for the surface textured seal [11] and the THD model for the spiral groove seal [12] were respectively investigated to illustrate the case of thermal instability. The performance of a textured face seal was compared with that of a smooth face seal by TEHD simulations and experiments, stressing that fluid-solid coupling must be considered when designing[13]. Meng et al. analyzed the thermal mixing effect induced by the U-shape notches on the seal face[14]. Song and Bai focused on the thermal cavitation effect of the liquid film seal at high-speed condition[15], while Ma and Li actively utilized the cavitation effect to suppress the temperature increasing[16].

The cited studies above primarily focused on steady-state behavior. However, seals often operate under transient conditions. Research into transient behavior, including deformation effects, has centered mainly on initially flat face seals. Harp and Salant [17] developed a mathematical model for the transient features of the hydrostatic face seals, including the contact mechanics. Evolutions of typical parameters were demonstrated under varying speed and pressure. Through numerical simulations, Green [18] found that the time-dependent deformation is inherent and lags the instantaneous heat generation. The face-separating mechanism of the hydrostatic face seal during start-up was also emphasized. Danos et al. [19] developed a 2D transient THD model of an aligned face seal to analyze the viscous heat dissipation in the interface during start-up. This model was then improved by Tournerie et al. [20] by taking the thermo-elastic distortions into account, and a thermal instability state was simulated. Salant and Cao [21] proposed a semi-empirical approach to predict the transient performance of a contacting face seal, finding that the thermal lag and the squeeze film could lead to unsteady effects. Gao et al. [22] adapted a quasi-steady-state approach, which neglected the influence of thermal lag and inertia although considered the thermal deformation, to simulate the start-up process of a contacting face seal, finding that the coning direction of the seal gap changes accompanying by a decreasing contact area and an increasing leakage. Cochain et al. [23] developed a numerical model coupling a transient Reynolds equation, an analytical contact model, and solvers of force balance and thermo-mechanical deformations. Ran et al. [24] simultaneously considered the mechanical-thermal deformation and the dynamic response, pointing out significant difference comparing with simpler models at high-parameter ranges.

As investigated above, the dynamic behaviors and the thermal features are both inherent attributes of mechanical face seals, and they are strongly coupled with and interactively influenced by each other. Due to the dramatical complexity caused by the different time scales, either the dynamics or the TEHD lubrication is only individually considered when modeling. Several studies aiming at the transient TEHD lubrication characteristics did assumptions such as 1D fluid field, ignoring the dynamic equations, quasi-steady-state during varying operating conditions, etc., making it more practicable but still restricted.

In the present work, a comprehensive numerical solution is established for the transient mixed TEHD lubrication of mechanical face seals, coupling the transient Reynolds equation, the transient energy equation, time-dependent heat conduction and thermal deformation models for the solid, dynamic equations of the flexibly mounted seal ring, equation of the sealing gap clearance, as well as rough face contact model. To make the solution executable in spite of very different time scales of the dynamic and thermal features, an approach of Parallel Dual Time Steps (PDTS approach) is proposed. Both of the efficiency and accuracy of the PDTS approach are evaluated by comparing with the reference. Furthermore, an outer-herringbone-grooved face seal in a start-up stage is studied. The primary objective is to elucidate the contributions of distinct physical phenomena. Transient stages like start-up involve time-dependent evolutions of thermal effects and dynamic displacements that govern seal performance. Demonstrating these evolutionary processes and revealing their influence will inform both seal design and failure mechanism analysis.

## 2. Studied configuration

The target mechanical face seal is designed based on a former test cell as shown in **Fig. 1(a)** [13]. The structure of the seal assembly is illustrated in **Fig. 1(b)**. The stator is a transparent ring made of sapphire, and the rotor is flexibly mounted and thrusted by cylinder springs. The herringbone grooves are configured on the rotor face with outer-half grooves connecting the pressured water. The structural parameters are marked in **Fig. 1(c)**, and the radial widths of the dam, the inner-half grooves and the outer-half grooves are respectively 2.2 mm, 2.3 mm and 1.1 mm.

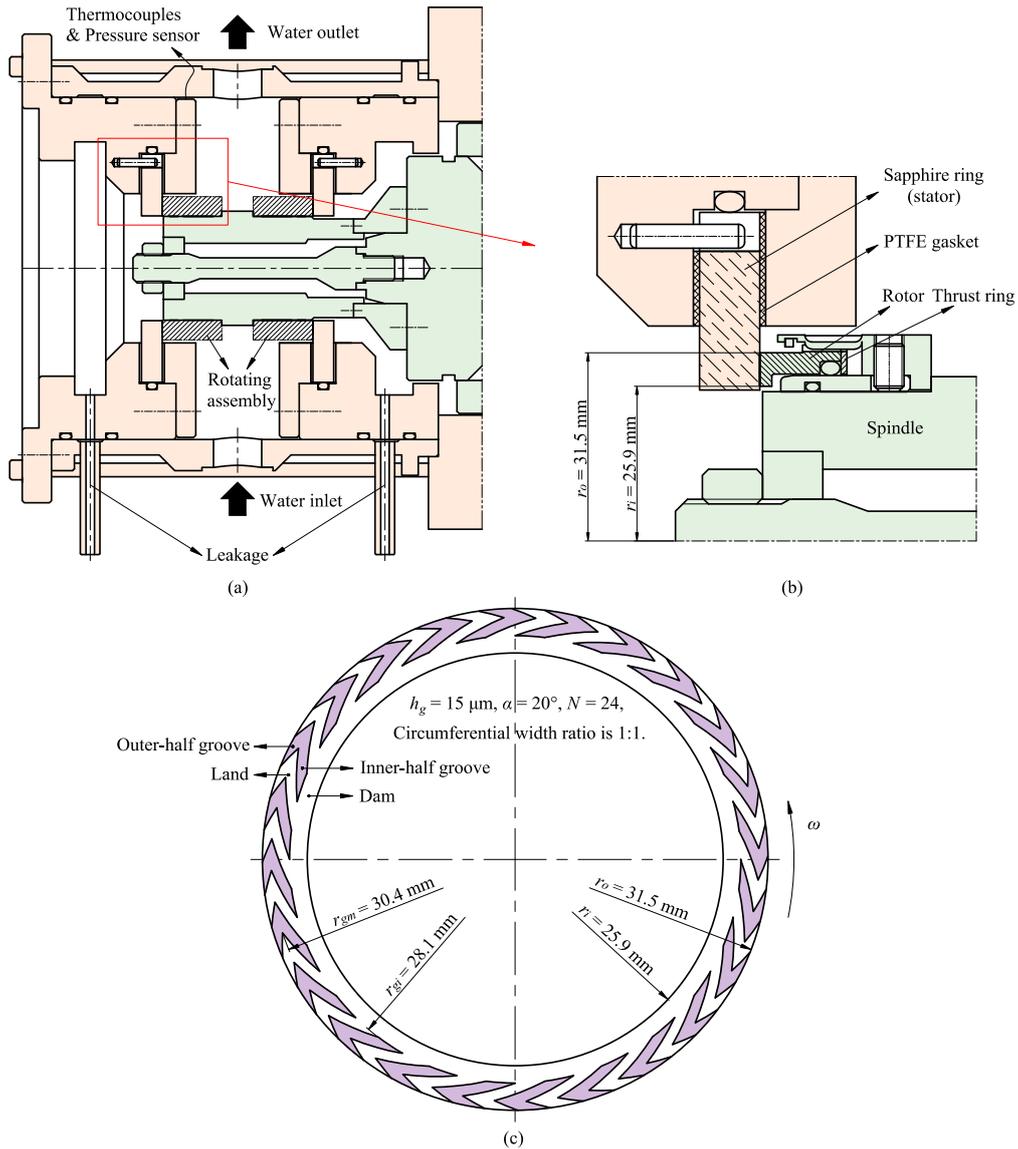

**Fig. 1 Structure of the target face seal: (a) test cell, (b) assembly of the face seal, (c) herringbone groove structure on the rotor face.**

## 3. Mathematical models

Several physical effects and mechanisms simultaneously work during running of a face seal, which can be summarized as:

(1) hydrodynamic and/or hydrostatic lubrication of the fluid film;

(2) the fluid film rheology;

(3) heat transfer among the rotor, the stator, the fluid film, the medium and the ambient;

(4) mechanical and thermal deformations (especially the axial deforming) of the rotor and stator faces governed by the solid mechanics;

(5) multi-DOF (Degrees of Freedom) dynamics of the system composed of the flexibly mounted seal

ring, the rotor, the supporting elements (e.g., springs and O-ring) and the fluid film;

(6) contacting and friction (even wear) of the rotor and stator faces.

In order to establish the transient mixed TEHD lubrication theoretical model as well as the numerical model for the solution, the elements mentioned above are all involved in the mathematical models.

### 3.1. Reynolds equation and cavitation effect model

The Reynolds equation including the average flow model is adapted to govern the hydrodynamic lubrication of the fluid film in the sealing gap. It is assumed that the volume force and the inertial effect, the wall slip, the pressure and temperature variations through the film thickness, and the turbulence effect are all ignored. The Reynolds equation in polar coordinates is given as:

$$\frac{1}{r}\frac{\partial}{\partial r}\left(r\phi_p \frac{\rho h^3}{12\mu}\frac{\partial p}{\partial r}\right) + \frac{1}{r}\frac{\partial}{\partial \theta}\left(\frac{1}{r}\phi_p \frac{\rho h^3}{12\mu}\frac{\partial p}{\partial \theta}\right) = \frac{1}{r}\frac{\partial}{\partial \theta}\left[\rho \frac{r\omega}{2}\left(\bar{h}_T + \sigma\phi_{s,rs}\right)\right] + \frac{\partial \left(\rho \bar{h}_T\right)}{\partial t} \quad (1)$$

in which the expressions of the pressure flow factor $\phi_p$ and the shear flow factor $\phi_{s,rs}$ are based on the isotropic surface assumption proposed by Patir and Cheng [25][26].

The general model based the homogeneous flow hypothesis proposed by Brunetiere [27] is utilized to describe the cavitation effect in the liquid film. Core assumptions of this general model are stated as:

(1) the fluid is a homogeneous mixture of an incompressible liquid of density $\rho_l$ and a perfect gas of density $\rho_g$;

(2) the surface tension at the liquid-gas interface is neglected, therefore the local gas pressure is equal to the local liquid pressure;

(3) two phases move at the same velocity;

(4) the mass fraction $\lambda_g$ of gas is uniform in the whole fluid domain, and the liquid and the gas are water and air in this work where $\lambda_g$ is selected as 0.000075 [27].

In this proposed theory, the density of the homogeneous fluid is expressed as:

$$\rho = \rho_M \frac{p}{p + p_t} \quad (2)$$

in which the theoretical maximum density is $\rho_M = \rho_l/(1 - \lambda_g)$; the pressure at half $\rho_M$ (called transition pressure) is $p_t = \lambda_g \rho_l r_{gas}/(1 - \lambda_g) \cdot T$.

The gas volume fraction representing the cavitation effect is calculated by:

$$\alpha = \frac{p_t}{p + p_t} \quad (3)$$

*3.2. Energy equation*

Temperature of the fluid film is governed by the energy equation. Besides constant axial pressure and temperature, two more assumptions are: i. only the shear strain rates $\partial u/\partial z$ and $\partial v/\partial z$ are considered; ii. the thermal conductance along *r*-direction and *θ*-direction in the fluid film is ignored. The energy equation in polar coordinates is given as [28]:

$$\lambda(T_r + T_s - 2T) + \frac{\mu}{h}\omega^2 r^2 + h\frac{\partial p}{\partial t} + q_r\frac{\partial p}{\partial r} + q_\theta\frac{1}{r}\frac{\partial p}{\partial \theta} = \rho h\frac{\partial(c_p T)}{\partial t} + \rho q_r\frac{\partial(c_p T)}{\partial r} + \rho q_\theta\frac{1}{r}\frac{\partial(c_p T)}{\partial \theta} \tag{4}$$

The effect of convective heat transfer between the fluid film and the faces is based on the Chilton and Colburn Analogy [29], and the convective coefficient in the sealing gap is given as:

$$\lambda = \Pr^{\frac{1}{3}} \cdot \frac{\rho \bar{v}^2 k}{ph} \tag{5}$$

in which $\Pr = \mu c_p/k$ is the Prandtl number of the fluid film; $\bar{v}$ is the average circumferential velocity of the fluid film.

*3.3. Dynamic equation*

According to the Law of Momentum Conservation, the dynamic equation of the axial movement is given as:

$$M \cdot \ddot{z}_r + D \cdot \dot{z}_r + K \cdot z_r = F_o - F_c \tag{6a}$$

moreover, the dynamic equations for iteratively solving the displacement, the velocity and the acceleration of the reference (as shown in **Fig. 2**) of the flexibly mounted seal ring (it is the rotor here) are given as:

$$[z_r^{(k+1)}] = [z_r^{(k)}] + [\dot{z}_r^{(k)}] \cdot \Delta t^{(k)} + \frac{1}{2}[\ddot{z}_r^{(k)}] \cdot (\Delta t^{(k)})^2 \tag{6b}$$

$$[\dot{z}_r^{(k+1)}] = [\dot{z}_r^{(k)}] + [\ddot{z}_r^{(k)}] \cdot \Delta t^{(k)} \tag{6c}$$

$$[\ddot{z}_r^{(k+1)}] = [M^{(k+1)}]^{-1} \cdot \{[F^{(k+1)}] - [D^{(k+1)}] \cdot [\dot{z}_r^{(k+1)}] - [K^{(k+1)}] \cdot [z_r^{(k+1)}]\} \tag{6d}$$

in which (*k* + 1) represents the current time step in the temporal dimension, and (*k*) represents the former time step. Although the 3-DOF dynamic equations have been established, but only the axial movement of the flexibly mounted rotor is considered and presented in this work. Therefore, according to **Fig. 2**, the displacement, velocity and acceleration vectors $[z]$, $[\dot{z}]$ and $[\ddot{z}]$ are respectively the axial displacement $z$, velocity $\dot{z}$ and acceleration $\ddot{z}$ of the rotor face reference; the inertia, damping and stiffness matrices $[M]$, $[D]$ and [$K$] are respectively the mass $M$ of the rotor, and the damping $D$ and the stiffness $K$ of the supporting elements; the force vector [$F$] is the resultant force of the opening force and the closing force acting on the rotor, and it is $F = F_o - F_c$.

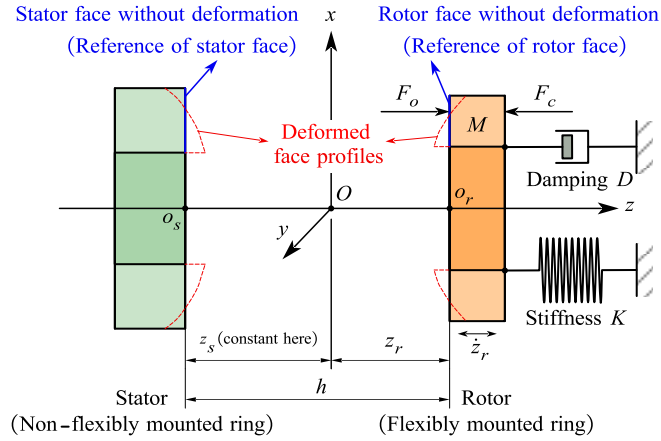

**Fig. 2 Kinematic model of the FMR face seal (axial movement).**

*3.4. Contact model*

The contacting pressure is calculated based on a statistical elastic-plastic contact model. The assumptions are consistent with those in the GW model [30] and the CEB model [31]. It is also assumed that two rough surfaces can be equivalent to one smooth and rigid surface and one rough surface. One single asperity pressed by a rigid surface is shown in **Fig. 3**. In the elastic deformation stage, the contacting pressure to clearance relationship is based on the Hertz contact theory, while in the elastic-plastic deformation stage, it is based on a general solution presented by Kogut and Etsion [32].

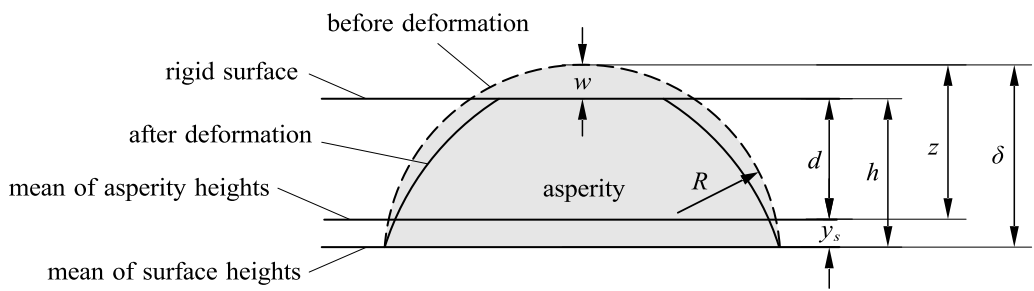

**Fig. 3 An asperity pressed by a rigid surface.**

A sample surface of the rotor face is measured by the profilometer, and the profile around its mean height level is shown as **Fig. 4** For the surface height distribution, the ratio of the arithmetical mean deviation $Ra$ and the standard deviation $Rq$ is 0.789, which is very close to the ratio of the Gaussian distribution 0.798. Therefore, assuming that the height distribution from the mean of asperity heights follows the Gaussian distribution, the probability density function can be given as:

$$g_s(z) = \frac{1}{\sigma_s \sqrt{2\pi}} \cdot e^{-\frac{z^2}{2\sigma_s^2}} \tag{7}$$

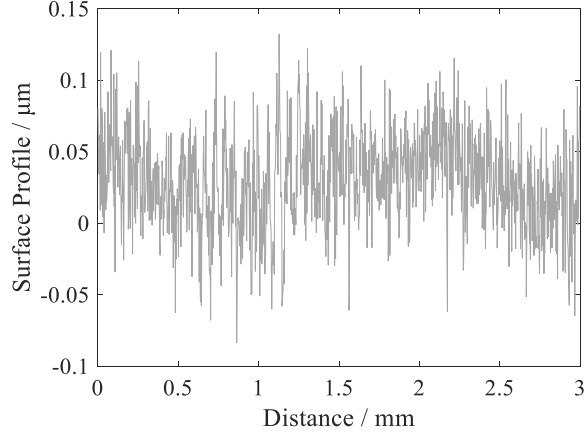

**Fig. 4 Sample profile of the rotor face.**

The macroscopic contacting pressure on the nominal contacting area can be calculated by the following integral formula [32]:

$$p_{ct} = \eta \cdot \left[ \int_d^{d+w_{cr}} P_{cr} \left(\frac{w}{w_{cr}}\right)^{1.5} g_s(z) \mathrm{d}z + \int_{d+w_{cr}}^{d+6w_{cr}} P_{cr} \times 1.03 \left(\frac{w}{w_{cr}}\right)^{1.425} g_s(z) \mathrm{d}z \right. \\ \left. + \int_{d+6w_{cr}}^{d+110w_{cr}} P_{cr} \times 1.40 \left(\frac{w}{w_c}\right)^{1.263} g_s(z) \mathrm{d}z + \int_{d+110w_{cr}}^{\infty} 2\pi RwH \cdot g_s(z) \mathrm{d}z \right] \tag{8}$$

and the detailed derivation can be also induced from this reference.

## 4. Methodology

### 4.1. Numerical model of fluid field

As shown in **Fig. 5**, the numerical computational domain of the fluid field is the thin fluid film ring (or one period of the ring) in the sealing gap, and the domain is discretized utilizing the Finite Volume Method (FVM). As shown in **Fig. 5(c)**, the control volume faces are cut by the groove boundaries, and the technique introduced by Kogure et al. [33] (called Kogure-tech) is implemented to accommodate these discontinuities. The groove segment length ratios $R_{grn}$, $R_{grs}$, $R_{gre}$ and $R_{grw}$, defined as ratios of the segment length in the groove and the whole length respectively for the north, south, east, west faces of each control volume in the FVM, as well as $R_{ldn} = 1 - R_{grn}$, $R_{lds} = 1 - R_{grs}$, $R_{lde} = 1 - R_{gre}$, $R_{ldw} = 1 - R_{grw}$ and $\bar{R}_{gr} = (R_{grn} + R_{grs} + R_{gre} + R_{grw})/4$, are used in discretizing **Eq. (1)** and **Eq. (4)**.

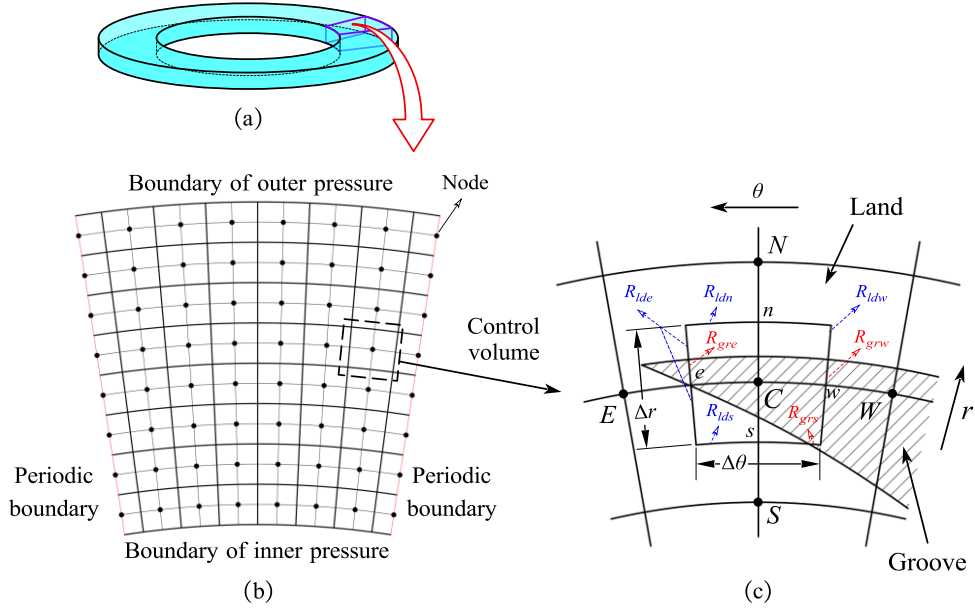

Fig. 5 Numerical discretion scheme of the fluid field: (a) fluid film, (b) single-period computational domain, (c) a control volume in the FVM.

The discretized Reynolds equation (**Eq. (1)**) based on the FVM and the Kogure-tech is derived as:

$$p_C = \frac{C_{pN} \cdot p_N + C_{pS} \cdot p_S + C_{pE} \cdot p_E + C_{pW} \cdot p_W - C_{pO}}{C_{pC}} \qquad (9)$$

in which the coefficients are presented in **Appendix A**.

On the basis of **Eq. (9)**, the iteration formula adapting the method of successive relaxation (SR) for solving the fluid film pressure is given as:

$$p_{C\,n+1}^{k+1} = (1-\omega_{rf,p}) \cdot p_{C\,n}^{k+1} + \omega_{rf,p} \cdot f\left(p_{C\,n}^{k+1}\right) \qquad (10)$$

in which $\omega_{rf,p}$ is the relaxation factor, and the value 1.1 is used here.

The discretized energy equation (**Eq. (4)**) based on the FVM and the Kogure-tech is derived as:

$$\begin{aligned}C_{TC}^{k+1} T_{C\,n+1}^{k+1} &= C_{TO}^{k+1} - C_{TN}^{k+1} \cdot T_{N\,n}^{k+1} + C_{TS}^{k+1} \cdot T_{S\,n}^{k+1} \\ &\quad - C_{TE}^{k+1} \cdot T_{E\,n}^{k+1} + C_{TW}^{k+1} \cdot T_{W\,n}^{k+1} + C_{TCk} \cdot T_C^k\end{aligned} \qquad (11)$$

in which the coefficients are presented in **Appendix B**.

On the basis of **Eq. (11)**, the iteration formula adapting the method of successive relaxation (SR) for solving the fluid film temperature is given as:

$$T_{C\,n+1}^{k+1} = (1-\omega_{rf,T}) \cdot T_{C\,n}^{k+1} + \omega_{rf,T} \cdot f\left(T_{C\,n}^{k+1}\right) \qquad (12)$$

in which $\omega_{rf,T}$ is the relaxation factor, and the value 1.1 is used here.

The convergent condition is expressed as:

$$\max_{\substack{1 \leq i \leq i_{\max} \\ 1 \leq j \leq j_{\max}}} \left| \frac{X_{i,j}^{(m+1)} - X_{i,j}^{(m)}}{X_{i,j}^{(m)}} \right| < 10^{-5} \tag{13}$$

in which $X_{i,j}^{(m)}$ represents the variable $X$ (e.g. $p$ or $T$) at the node $(i, j)$ and the iteration step $m$.

### 4.2. Fluid-solid coupling

As for the solid, the transient thermal conduction, the transient thermal deformation and the mechanical deformation of the seal rings are all involved in the transient mixed TEHD lubrication model. Particularly, the effect of thermal lag is included.

#### 4.2.1. 2D models and boundary conditions of the seal rings

In this methodology, the rotor and the stator are simplified into a 2D axisymmetric model. The mechanical boundaries and the thermal boundaries of the seal rings are presented in **Fig. 6**. In addition to the clear illustration in the figure, what should be further stated is:

(1) the rotor is held by the thrust ring at the back with the constraint type of frictionless contact;

(2) the stator is fixed by two PTFE gaskets with a certain compression of -20 μm, and the interfaces are also set as frictionless contact;

(3) for the rotor, two boundaries connecting and pressing the O-ring are set as the mechanical boundary type of medium pressure and the thermal boundary type of adiabatic.

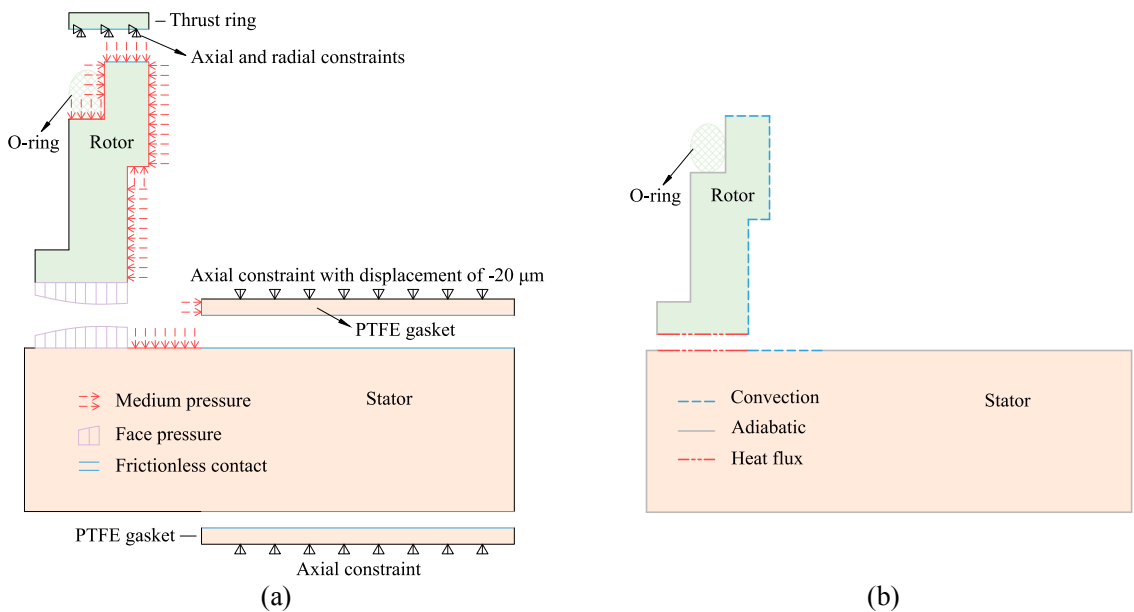

**Fig. 6 2D models and boundary conditions of the seal rings: (a) mechanical boundary conditions, (b) thermal boundary conditions.**

#### 4.2.2. Solving method of the solid features

The Influence Coefficient Method (ICM) is implemented to calculate the steady-state temperature and axial deformations of the faces. Particularly, the Duhamel's Principle is introduced to solve the transient temperature and the transient thermal deformation. The precondition to adapt both methods is that the objective system is a linear system. Furthermore, the Duhamel's summation formulas are given as [21]:

$$[\Delta T(t_s^{(l)})]_{n\times 1} = \sum_{m=1}^{l}\left\{[IC_T]_{n\times n}\cdot[H_s(\tau_s^{(m)})]_{n\times 1}\cdot\Delta\tau_s^{(m)}\cdot\bar{\tilde{S}}_{\Delta T}{'}(t_s^{(l)}-\tau_s^{(m)})\right\} \tag{14}$$

$$[u_{th}(t_s^{(l)})]_{n\times 1} = \sum_{m=1}^{l}\left\{[IC_{th}]_{n\times n}\cdot[H_s(\tau_s^{(m)})]_{n\times 1}\cdot\Delta\tau_s^{(m)}\cdot\bar{\tilde{S}}_{th}{'}(t_s^{(l)}-\tau_s^{(m)})\right\} \tag{15}$$

in which $[\Delta T(t_s^{(l)})]_{n\times 1}$ is the transient temperature increment of the face at $t_s^{(l)}$;

$[u_{th}(t_s^{(l)})]_{n\times 1}$ is the transient axial thermal deformation of the face at $t_s^{(l)}$;

$[IC_T]_{n\times n}$ is the influence coefficient matrix of the steady-state temperature of the face;

$[IC_{th}]_{n\times n}$ is the influence coefficient matrix of the steady-state axial thermal deformation of the face;

$[H_s]_{n\times 1}$ is the heat flux acting on the face;

$\tau_s^{(m)}$ is the dummy temporal variable;

$\bar{\tilde{S}}_{\Delta T}{'}$ is the mean of the step responses of temperature increment at all face nodes;

$\bar{\tilde{S}}_{th}{'}$ is the mean of the step responses of axial thermal deformation at all face nodes.

The influence coefficient matrices and the step responses are calculated by the software ANSYS Mechanical APDL based on the Finite Element Method (FEM).

*4.3. Nominal clearance of sealing gap*

As shown in **Fig. 2**, the nominal clearance of the sealing gap (also the nominal thickness of the fluid film) $h$ is related to the displacements of the rotor face reference and the stator face reference $z_r$ and $z_s$, and the clearance increment $\Delta h_u$ induced by face deformations. Besides, the depth of the face grooves $h_g$ should be considered as well. Therefore, the sealing gap clearance distribution can be given as:

$$h = z_r - z_s + \Delta h_u + \bar{R}_{gr}\cdot h_g \tag{16}$$

in which,

$$\Delta h_u = [(u_{me1,r}-u_{me1,s})+(u_{me2,r}-u_{me2,s})]+(u_{th,r}-u_{th,s}) \tag{17}$$

where $u_{me1,r}$ and $u_{me1,s}$ (the axial mechanical deformations of the rotor and stator faces caused by the face pressure - both fluid film pressure distribution and contacting pressure distribution are included), and $u_{me2,r}$ and $u_{me2,s}$ (the axial mechanical deformations of the rotor and stator faces caused by the medium pressure) are calculated by the ICM; $u_{th,r}$ and $u_{th,s}$ (the transient axial thermal deformations of the rotor and stator faces) are calculated by the Duhamel's Method.

*4.4. Iteration scheme of transient mixed TEHD lubrication solving*

The transient mixed TEHD lubrication of mechanical face seals are numerically solved by coupling several modules mentioned above via the temporal course. The flow chart of the iteration procedures is presented in **Fig. 7**. The constant parameters of the whole calculation are prescribed in the Parameter Inputting, defining several groups of properties as shown in the module, following which is the Initialization module where the initial values of the temporally varying parameters are input. Afterwards, the main iteration along temporal dimension starts. The time and the operating parameters are updated firstly. Then the displacement and the velocity of the FMR reference are calculated by **Eq. (6a)** and **Eq. (6b)**. Concurrently, the solid properties are calculated by the ICM and the Duhamel's Method. Then the nominal sealing gap clearance distribution can be obtained by **Eq. (16)**. Furthermore, the iterative formula **Eq. (10)** of the Reynolds equation and the iterative formula **Eq. (12)** of the energy equation are coupled for solving the fluid film properties. Parallel, the distribution of the contacting pressure is calculated according to the previously fitted function of the pressure-clearance relationship. The following part is "rotor face-fluid film-stator face" interface heat transfer in which the convective heat transfer between fluid film and solid surfaces, the thermal contact conductance between faces, as well as the contacting frictional power are all considered. The performance parameters of the current time step are then calculated. So far, the calculation of current time step is finished.

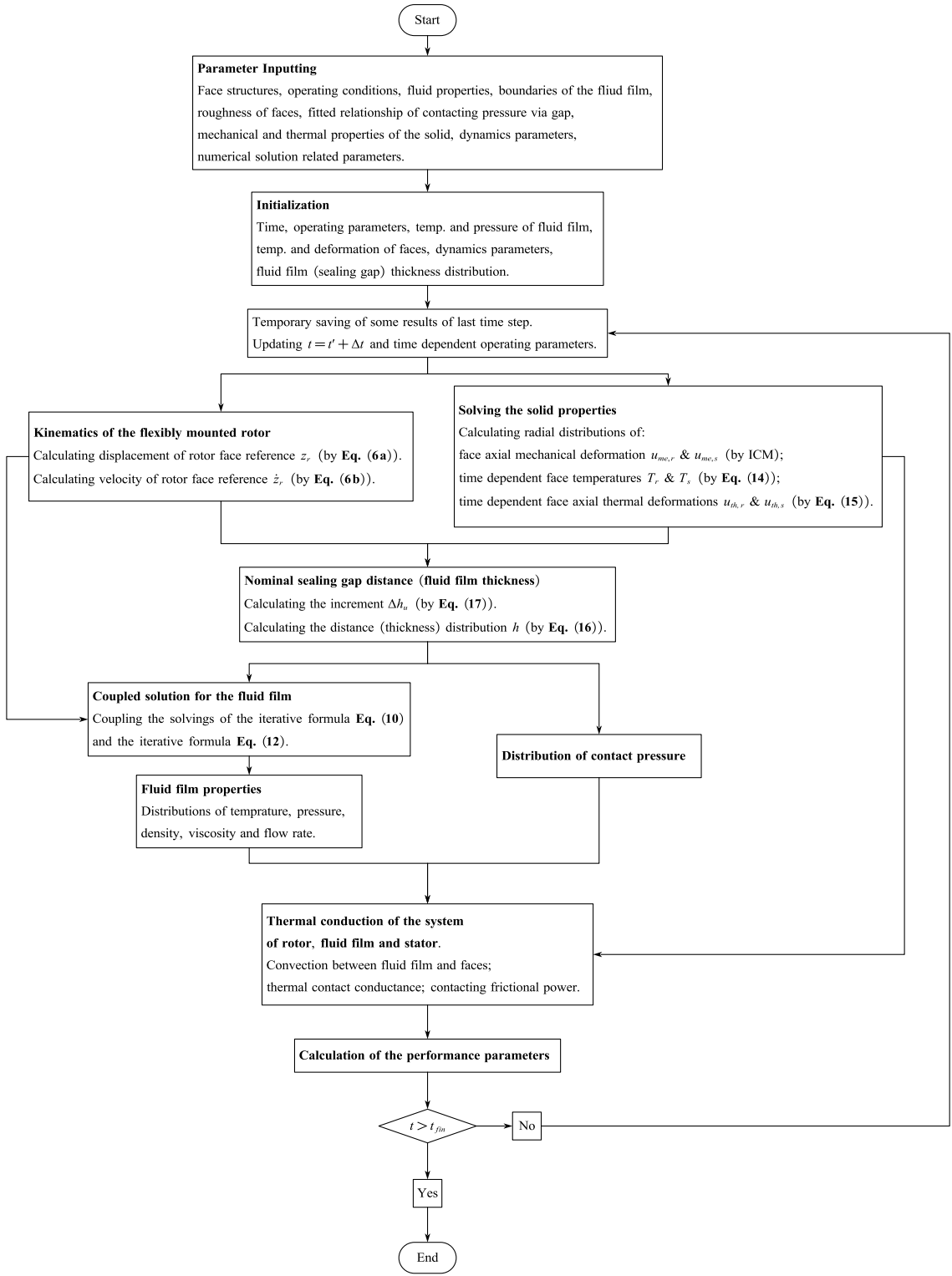

**Fig. 7 Flow chart of the main program of numerical calculation.**

Therefore, as shown in **Fig. 7** and stated above, the explicit time solver is adapted in the iterative calculation along the temporal dimension. Generally, problems could be met when coupling the transient thermal analyses and the dynamics for mechanical face seals due to very different time scales (second magnitude in the former and microsecond magnitude in the latter). Although the simulation can be executed with the explicit approach and very small time steps, it would be super costly in the calculation time. In order to resolve this problem, an approach of Parallel Dual Time Steps (PDTS approach) is proposed and utilized in the present work. In this approach, the transient solving of the solid properties is in a much lager time step of $10^{-4}$ s magnitude.

*4.5. Evaluation of the solution approaches*

To evaluate this comprehensive numerical solution and the PDTS approach, it should be benchmarked against prior studies. However, it is challenging to identify references that directly couple both transient processes. Consequently, a quasi-steady TEHD study [34] is selected to assess the efficiency and accuracy of the proposed method. Beyond the parameter configuration adopted from this reference, five additional rotational speed evolutions are introduced, as shown in **Fig. 8(a)**. All profiles converge to a final speed of 2000 r/min, matching the reference study. However, their transient trajectories differ significantly, with intermediate speeds ranging from 1000 r/min to 5000 r/min.

It should first be noted that all five solutions can be computed within one hour using a desktop computer with an 11th Gen Intel® Core™ i7-1165G7 processor. This computational efficiency enables comprehensive acquisition of progressive data. **Fig. 8(b)** displays evolutions of the maximum temperature increment at the rotor face (typically at the inner diameter). While significantly influenced by rotational speed, the temperature responses exhibit a delay relative to the speed changes. For instance, although all speed profiles reach 2000 r/min at t = 0.3 s, corresponding temperature increments vary substantially (51.62°C to 56.51°C, as shown in **Fig. 8(c)**), compared to the steady-state value of 53.50°C at 2000 r/min. This clearly demonstrates that transient thermal behavior depends not only on current operating conditions but also on historical operational trajectories, demonstrating the effect of the thermal lag.

The face clearance profile and fluid film temperature distribution are shown in **Fig. 8(d)**. Compared with the reference results (as shown in **Fig. 8(e)**), several key consistencies emerge. First, a convergent clearance along the leakage direction is achieved, primarily due to thermal deformation of the faces. Second, at any given radial position, the film temperature exceeds that of both the rotor and stator faces, demonstrating that heat is generated within the film and dissipates to the faces. Third, the minimum film thickness is significantly larger than that in the mixed lubrication regimes, confirming the generation of hydrostatic load-carrying capacity. This alignment with the reference data provides strong evidence for the accuracy of the results.

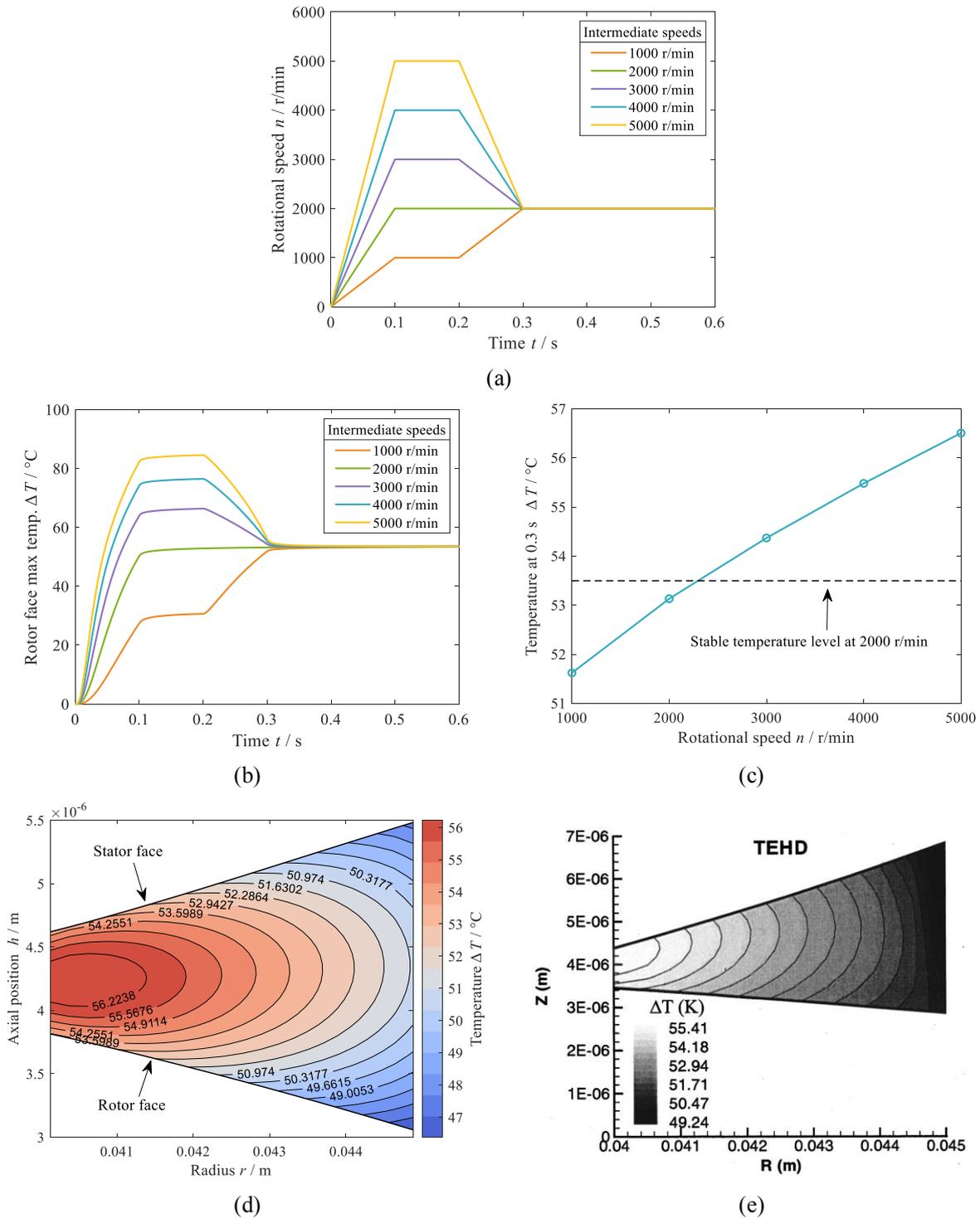

**Fig. 8** Evaluation of the solution approaches: (a) five introduced rotational speed evolutions, (b) evolutions of maximum temperature increment at rotor face, (c) comparation of temperatures at 0.3 s, (d) face clearance and film temperature profile at 0.6 s and 2000 r/min (current study), (e) face clearance and film temperature profile (from the reference [34])

## 5. Results and discussion

### 5.1 Parameters

The operating parameters used in simulations are set as same as those in the tests. The experimental apparatus used in the present work has been introduced in published studies [13]. The real-time data of the rotation speed and the face frictional torque are monitored and saved via a self-developed data acquisition software based on the LabVIEW platform. The front view of the test cell with outer-herringbone-grooved rotor assembled is shown in **Fig. 9**.

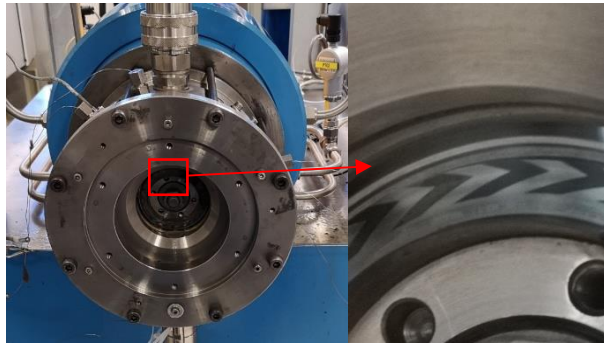

**Fig. 9 Front view of the test cell.**

#### 5.1.1. Data collection

The involved parameters are listed in **Table 1**. The pressures and the temperatures are constant, and the evolution of the rotational speed is illustrated in **Fig. 10**. From 0 s to 2.67 s, the speed rises from 0 r/min to 3405 r/min in about S-shape form which is fitted based on the test data. Following the speed-up stage is the steady-state period at 3405 r/min until 8 s.

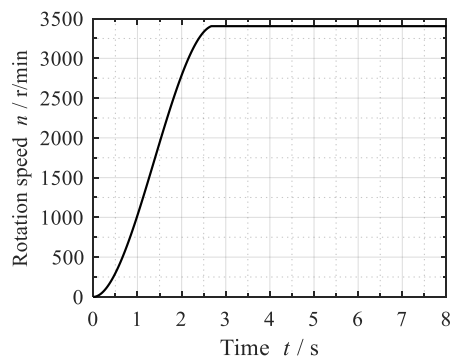

**Fig. 10 Evolution of the rotation speed.**

**Table 1 Collection of involved parameters**

| Parameter | Value | Parameter | Value |
| --- | --- | --- | --- |
| Pressured medium | Water | Rotor material | Tungsten carbide with Cobalt (WC-Co) |
| Ambient pressure | 0.1 | Stator material | Sapphire |

| Parameter | Value | Parameter | Value |
|---|---|---|---|
| (inner diameter) $p_i$ / MPa | | | |
| Medium pressure (outer diameter) $p_o$ / MPa | 0.58 | Dry frictional coefficient of the paired materials $\mu_{ct}$ | 0.1, 0.07, 0.04, 0.03 |
| Ambient temperature $T_i$ / °C | 25 | Face roughness $\delta$ / μm | 0.1 (stator face) 0.2 (non-groove area of rotor face) 0.8 (grooves of rotor face) |
| Medium temperature $T_o$ / °C | 60 | Thermal conductivity $k$ / W/(m·K) | 60 (WC-Co), 35 (Sapphire) |
| Balance ratio of the face seal $B$ | 0.7 | Stiffness coefficient of the supporting elements $K$ / N/m | 36400 |
| Spring force $F_{sp}$ / N | 116.2 | Damping coefficient of the supporting elements $D$ / N·s/m | 1200 |
| | | Mass of the flexibly mounted rotor $M$ / kg | 0.140 |

In this numerical solution, the water density is set as 1000 kg/m³, and the specific heat capacity is set as 4186 J/(kg·°C). The viscosity-temperature and conductivity-temperature relationships of the water are respectively given as:

$$\mu = \frac{1}{0.12087 \times (T+81.066)^2 - 236.5} + 1.3448 \times 10^{-7} \cdot T \quad (18)$$

$$k = \frac{0.56441 + 0.0019131 \cdot T}{1 + 6.5982 \times 10^{-6} \cdot T^{2.0901}} \quad (19)$$

*5.1.2. Convective coefficients between surfaces and medium*

Based on a commonly recognized empirical formula for the convective coefficient of heat transfer at the interface of a rotating cylindrical surface and the fluid surrounding it [35], the convective coefficient of the rotor outer surface can be calculated under the parameters shown in **Fig. 1** and **Table. 1**. In addition, as one of the dependent variables, the angular velocity is varying as shown in **Fig. 10**. However, considering a short duration (several seconds) of acceleration, it is assumed that this convective coefficient is constant, and the empirical value at the maximum rotation speed is adopted. The convective coefficient between the rotor surface and the medium and that between the stator surface and the medium are separately 27000 W/(m²·°C) and 13500 W/(m²·°C).

*5.1.3. Parameter settings of simulation*

Ten groups of simulation are presented as shown in **Table 2**. Group A to C are based on different grid amounts of 50×50, 40×40 and 30×30 with the time step of 2×10⁻⁷ s, while group D, B and E are for time steps of $1\times10^{-7}$ s, $2\times10^{-7}$ s, $4\times10^{-7}$ s with the grid amount of 40×40. In order to compare with the TEHD case (Group B), Group F and G are respectively simulated with the THD and HD models, both with 40×40 and $2\times10^{-7}$ s setting. Three more groups (Group H, I and J) are conducted with lower dry frictional coefficient of the paired materials so as to comparing with the experimental results in company with Group B, F and G.

Table 2 Ten groups of simulation

| Group | Theoretical Model | Grid amount | Time step | Dry frictional coefficient of the paired materials |
|---|---|---|---|---|
| A | transient TEHD lubrication | **50×50** | $2\times10^{-7}$ s | 0.1 |
| **B** | **transient TEHD lubrication** | **40×40** | **$2\times10^{-7}$ s** | **0.1** |
| C | transient TEHD lubrication | **30×30** | $2\times10^{-7}$ s | 0.1 |
| D | transient TEHD lubrication | 40×40 | **$1\times10^{-7}$ s** | 0.1 |
| E | transient TEHD lubrication | 40×40 | **$4\times10^{-7}$ s** | 0.1 |
| F | **transient THD lubrication** | 40×40 | $2\times10^{-7}$ s | 0.1 |
| G | **transient HD lubrication** | 40×40 | $2\times10^{-7}$ s | 0.1 |
| H | transient TEHD lubrication | 40×40 | $2\times10^{-7}$ s | **0.07** |
| I | transient TEHD lubrication | 40×40 | $2\times10^{-7}$ s | **0.04** |
| J | transient TEHD lubrication | 40×40 | $2\times10^{-7}$ s | **0.03** |

*5.2. Independence evaluation of grid amount and time step*

To evaluate the influence of the grid scale, as well as the time step, on simulation results, the fluid force and the contacting force are compared among different groups. As shown in **Fig. 11(a)**, the load-carrying capacity generated by the fluid film (the fluid force) is first large since grooves connect the pressured water; afterwards, low-pressure areas in the film develop at groove roots with speed increasing, resulting in fluid force falling. With the mechanism of axial force balance, the contacting forces vary in the inverse tendency (as shown in **Fig. 11(b)**). Note that the band of these curves extends with width of around 5 N which is relatively limited. Gaps among curves of Group A, B and C (different gird amounts) take place at standstill and maintain along the process; while for Group B, D and E (different time steps) gaps mainly occur during speeding up. Considering the universal principle of balancing the accuracy of results and the efficiency of calculations, the moderate configuration is selected for the rest simulations, which is 40×40 grid scale and $2\times10^{-7}$ s time increment.

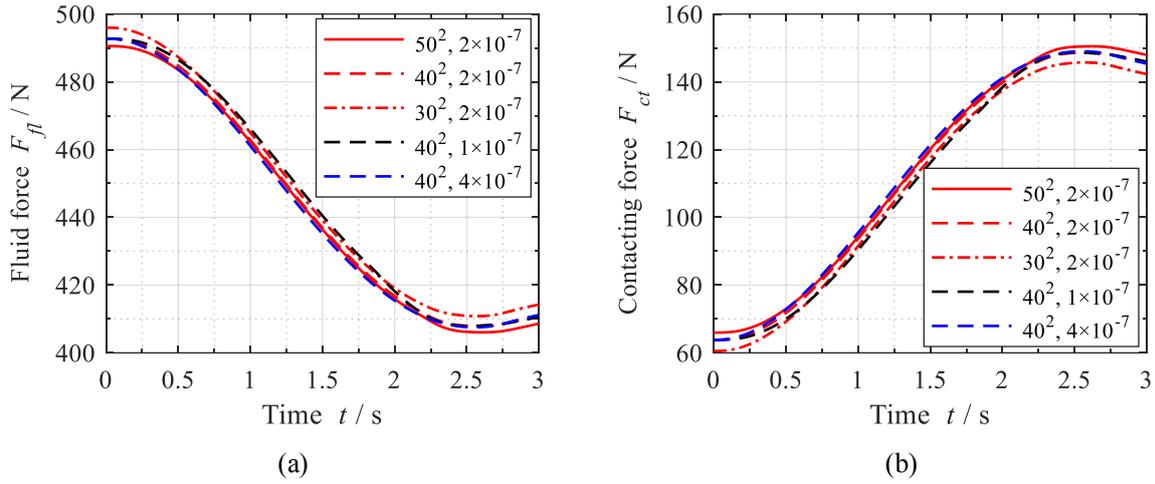

(a)  (b)

**Fig. 11 Independence evaluation of grid amount and time step: (a) fluid force, (b) contacting force.**

*5.3. Evaluation of the numerical model reliability*

    Theoretical results are compared with the experimental data so that the reliability of the numerical models can be examined and evaluated. Here the total frictional torques of Group B, F, G, H, I, J in **Table 2** and those of three repeated tests are presented and compared in **Fig. 12**. Note that a clear turning point occurs in all three curves of the test results due to the possibility that effective measurement of the data acquisition system will not start until around 1 s. It can be found that the theoretical torques are directly influenced by the dry frictional coefficient $\mu_{ct}$ utilized in the model. In the given cases, the results with $\mu_{ct}$ = 0.03 is most consistent with the measured curves. An unreasonable frictional coefficient in the theoretical model can overestimate the contacting frictional torque, however, the established numerical model is still sophisticated enough for exploring the transient mixed TEHD lubrication mechanism of face seals.

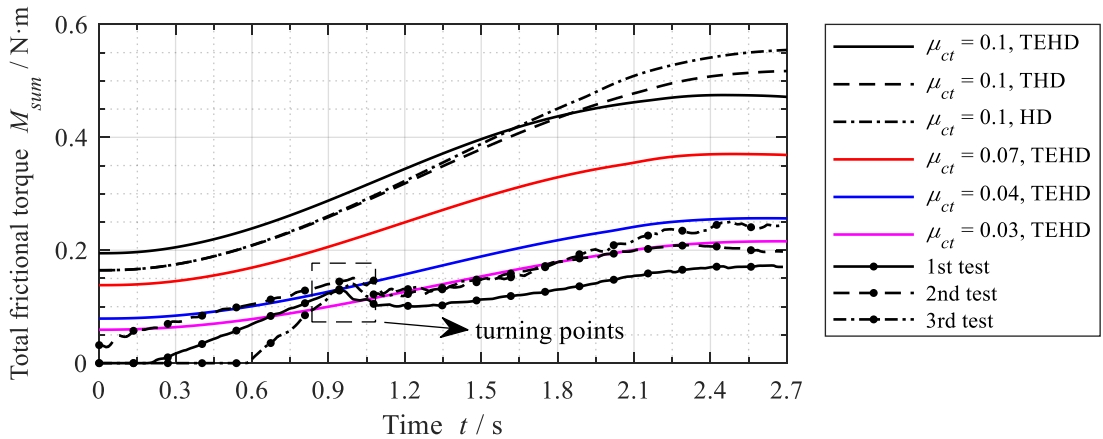

**Fig. 12 Comparison among simulations and tests.**

*5.4. Comparison among different models (TEHD, THD and HD)*

    To evaluate the influence of the thermal effect and the solid deformation on the theoretical results, the

simulation of the transient mixed THD lubrication (the THD case, Group F in **Table 2**) that ignores the solid deformation and the simulation of the transient mixed HD lubrication (the HD case, Group G in **Table 2**) that ignores both heat conduction and face deforming are illustrated together with the simulation of Group B (the TEHD case). Note that $\mu_{ct}$ in all of Group B, F and G are 0.1.

*5.4.1. Movement and deformation coupling properties*

Because the dynamic equations are contained in the transient TEHD, THD and HD models, the movement (only the axial movement is presented in this work) of the flexibly mounted rotor can be calculated.

Evolutions of the rotor reference displacements $z_r$ in three cases, as well as evolutions of the maximum and minimum fluid film thicknesses $h_{max}$ and $h_{min}$ in the TEHD case, are compared in **Fig. 13(a)**. It can be found that $z_r$ of the THD and HD cases are very close to each other and are relatively stable, but $z_r$ of the TEHD case is totally different from them and is in much larger variation. Actually, without the deformation of the seal rings, $z_r$ is same as the fluid film thickness of the non-groove area (when $z_s$ is zero). Therefore, $z_r$ of the THD and HD cases diminish a bit during acceleration to balance the closing force and then get steady.

As for $z_r$ of the TEHD case, it is smaller than $h_{min}$ at the beginning due to the mechanical deformation of the faces towards the negative *z*-direction. Afterwards, it starts to dramatically increase from 0.7 μm at 1 s to about 1.6 μm at 3 s and finally to 1.9 μm at 8 s. Note that $z_r$ is close to the evolution of $h_{max}$ in the TEHD case after 2 s, while the evolution of $h_{min}$ is obviously smaller than $z_r$.

Evolutions of the rotor reference velocities $\dot{z}_r$ in three cases are given in **Fig. 13(b)**. It can be found that both the magnitude and the direction of the velocities in the TEHD case and in the THD and HD cases are apart from each other. But all of three curves start from 0 m/s and go back to 0 m/s in the end.

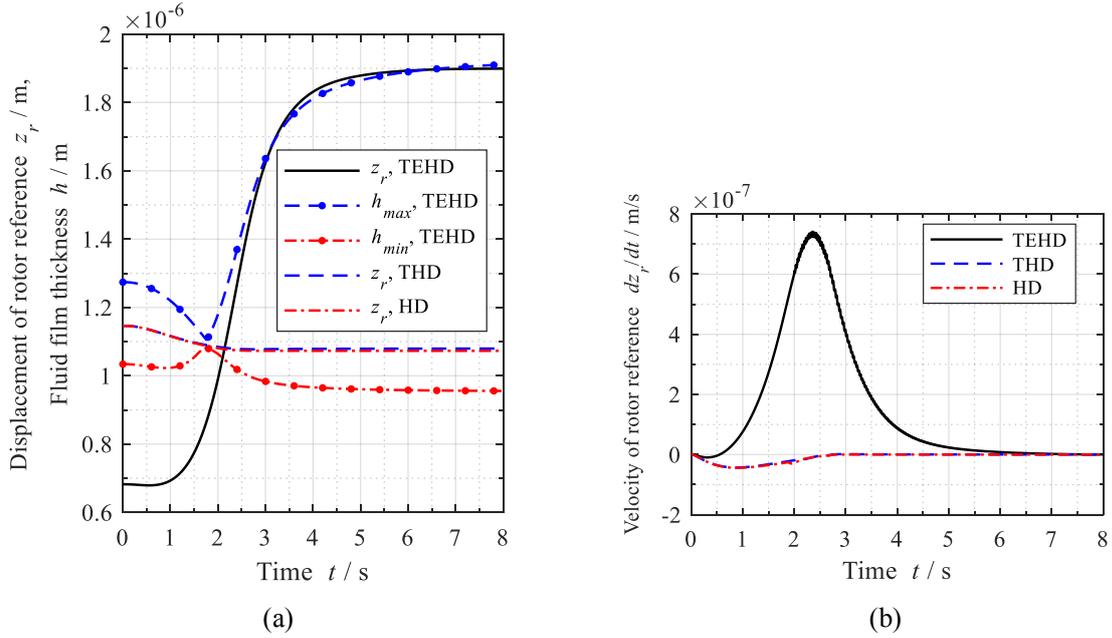

**Fig. 13** Movements of the flexibly mounted rotor: (a) displacements, (b) velocities.

Because the full fluid film lubrication fails to form (the contacting force always exists and even gets larger) in the given TEHD case, it can be concluded that the movements of the rotor reference towards the positive $z$-direction is induced mainly by the thrusting of the thermal expansion of the seal faces. Coupling of the movement and the deformation is successfully given.

*5.4.2. Contacting and friction properties*

Evolutions of the forces are presented in **Fig. 14(a)**. The fluid forces $F_{fl}$, the contacting forces $F_{ct}$ and the opening forces $F_o$ of the TEHD, THD and HD cases are compared. The overall tendencies of $F_{fl}$ and $F_{ct}$ are opposite to each other because $F_o$ is almost equal to the closing force which is constant, because the acceleration of the flexibly mounted rotor is super small. What is also common for three cases is that $F_{fl}$ decreases with speed rising, which is inconsistent with the empirical cognition and is discussed as the following.

The evolution of the fluid film pressure distribution in the TEHD case from 0 s to 3 s is illustrated in **Fig. 14(b)**. Because the grooves are connecting the pressured medium at the outer diameter, the liquid can invade the sealing gap even at downtime. At $t$ = 0 s, the film pressure at the grooves and the lands is almost uniform and equal to the sealed pressure, and it decreases to the inner pressure linearly along the dam. That is why the overall hydrostatic load-carrying capacity is enhanced at standstill. With rotation speed rising, the pressure diminishes at roots of the inner-half grooves. Even though high pressure emerges at joints of the inner-half and outer-half grooves, the fluid force is still weakened along the accelerating process. However, it does not mean that the outer herringbone grooves cannot improve the load-carrying capacity of the liquid film in the given operating condition, because a linear radial profile of the pressure from 0.58 MPa to 0.1 MPa can only generate the fluid force of 353.4 N which is obviously smaller than the values presented in **Fig. 14(a)**.

As shown in **Fig. 14(a)**, the fluid force $F_{fl}$ of the TEHD case is 6.6 N smaller than that of the THD and HD cases at standstill. This difference can be attributed to the slightly divergent sealing gap that is caused by the mechanical deformation and weakens the hydrostatic pressure. When it just reaches the maximum speed, the force values become lowest, respectively 407.9 N, 402.1 N and 391.8 N for the TEHD, THD and HD cases. Afterwards, the first two of them smoothly increase to 420.0 N and 406.7 N, while the last one stays stable. Therefore, the fluid force can be underestimated, meaning an overestimated contacting force, when the heat conductance and the solid deformation are neglected. Particularly, the variation of the fluid force at stable running cannot be given if the thermal lag effect is overlooked in the model.

Evolutions of the frictional torques are presented in **Fig. 14(c)**. The fluid, contacting and total frictional torques $M_{fl}$, $M_{ct}$ and $M_{sum}$ of the TEHD, THD and HD cases are compared. For the given cases, with $\mu_{ct} = 0.1$, the contacting torque is the dominant component of the total torque, showing that the overestimation of the torque compared to experiments is certainly due to an overestimation of the dry frictional coefficient. For the whole period of 8 s, the trend of the contacting torque is consistent with that of the contacting force, for example, $M_{ct}$ of the TEHD case goes back from the peak 0.407 N·m to nearly stable value 0.367 N·m, happening during the last five seconds.

As shown in **Fig. 14(c)**, $M_{fl}$ goes up with the speed rising, and slight differences among three cases smoothly appear during steady running. $M_{fl}$ at 8 s are respectively 0.058 N·m, 0.070 N·m and 0.082 N·m. $M_{sum}$ at 8 s are respectively 0.425 N·m, 0.501 N·m and 0.555 N·m, showing relatively larger differences. Therefore, ignoring the viscosity-temperature effect and the solid deformation can overestimate both contacting frictional torque and fluid frictional torque.

Evolutions of the contacting pressure distributions of the TEHD and THD cases from 0 s to 3 s are illustrated in **Fig. 14(d)**, and the maximum contacting pressure $p_{ct\ max}$ of three cases are compared in **Fig. 14(e)**. From the figures it can be seen that under the TEHD model, from 0 s to about 0.9 s, $p_{ct\ max}$ appears at the outer diameter and rises a bit; from about 0.9 s to about 1.8 s, $p_{ct\ max}$ obviously goes down to a valley value, revealing that the sealing gap transforms from slight divergence to parallel; after about 1.8 s, $p_{ct\ max}$ significantly climbs to about 0.9 MPa at 3 s from less than 0.3 MPa, following which is a continuously rising to more than 1.2 MPa, meanwhile, the contacting area moves towards the inner side of the face. It should be noted that although the contacting frictional torque diminishes after it reaches the maximum speed (as shown in **Fig. 14(c)**), the maximum contacting pressure keeps rising dramatically (as shown in **Fig. 14(e)**). As the contacting load per unit area is heavy enough, undesirable wear would occur on the faces. Therefore, the transient thermal deformation is directly related to the wear characteristics of the faces.

Under the THD and HD models, the contacting pressure is uniform on the non-groove area of the face and is significantly smaller than $p_{ct\,max}$ of the TEHD model. For instance, $p_{ct\,max}$ of three cases at 8 s are respectively 1.27 MPa, 0.236 MPa and 0.259 MPa. Consequently, the maximum contacting pressure is significantly underestimated and the pressure distribution is not realistic in the case of neglecting the face deformation.

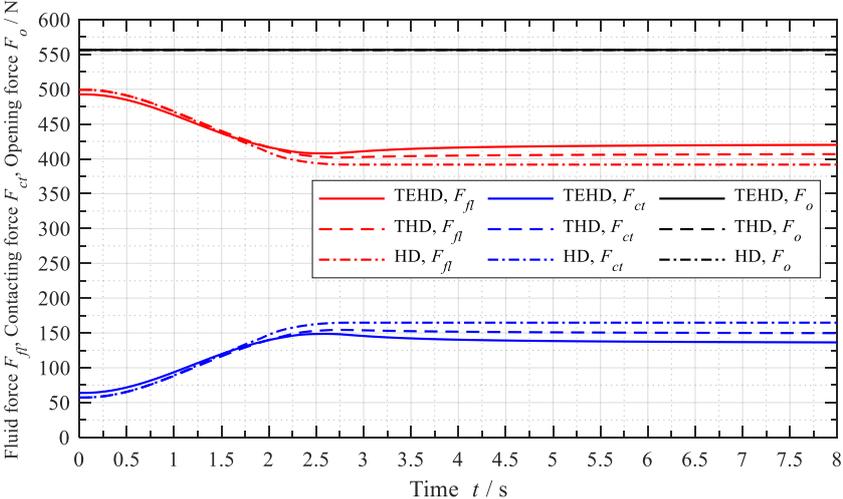

(a)

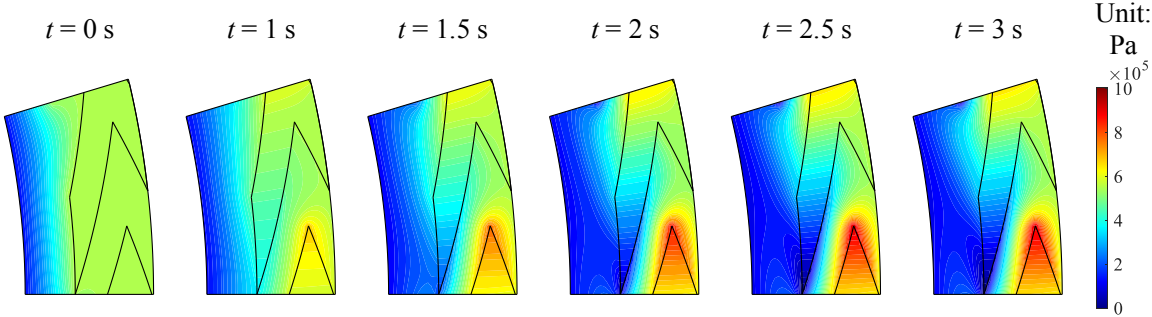

(b)

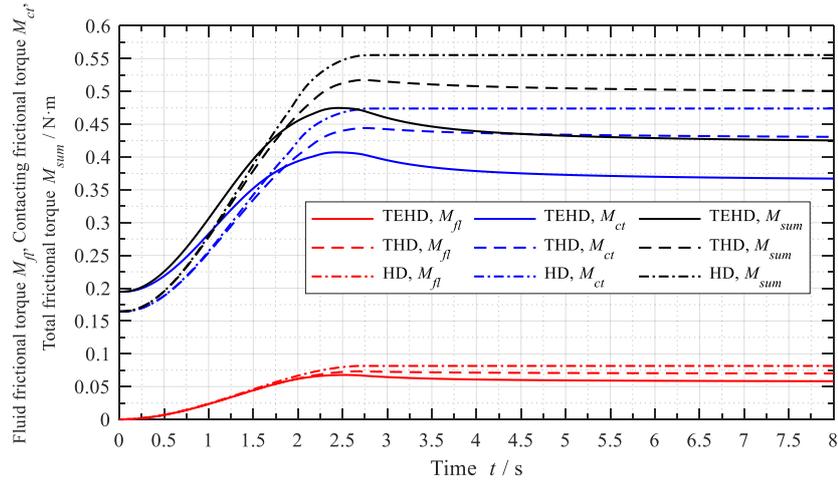

(c)

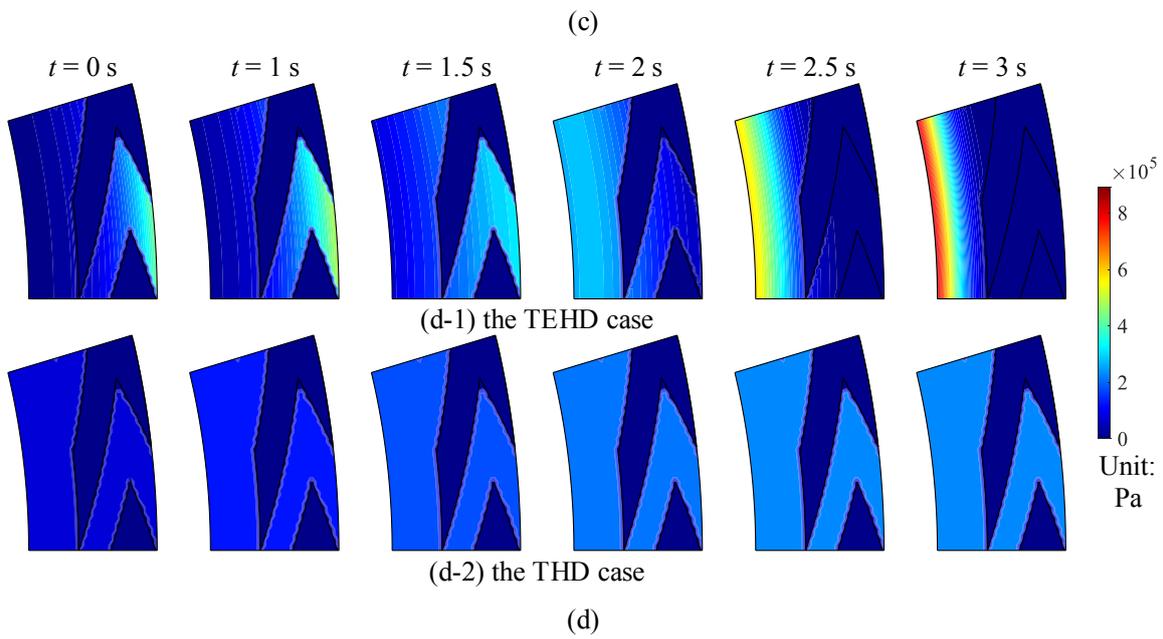

(d-1) the TEHD case

(d-2) the THD case

(d)

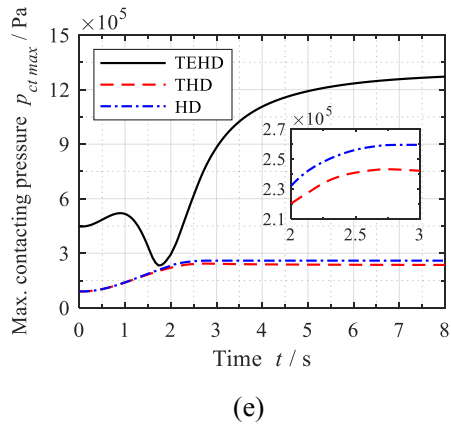

(e)

**Fig. 14** Contacting and friction properties: (a) evolutions of fluid forces, contacting forces and opening forces, (b) evolution of fluid film pressure distribution in the TEHD case (from 0 s to 3 s), (c) evolutions of fluid, contacting and total frictional torques, (d) evolutions of the contacting pressure distribution (from 0 s to 3 s), (e) evolutions of the maximum contacting pressures.

*5.4.3. Thermal properties*

Evolutions of the frictional power consumptions are presented in **Fig. 15(a)**. The thermal dissipations of the fluid film $P_{fl}$, the contacting frictional powers $P_{ct}$ and the total frictional powers $P_{sum}$ of the TEHD, THD and HD cases are compared. The overall tendencies of these powers are consistent with those of the frictional torques, and $P_{ct}$ is the main part of the total consumption. $P_{fl}$ of three cases at 8 s are separately 20.8 W, 25.0 W and 29.1 W, and $P_{sum}$ of three cases at 8 s are separately 151.7 W, 178.6 W and 198.1 W. The final $P_{sum}$ of the HD case is nearly one forth more than that of the TEHD case, and the final $P_{sum}$ of the THD case is nearly one fifth more than that of the TEHD case.

To compare with the radial temperature profiles of the rotor and stator faces, which are based on the 2D axisymmetric models, the temperature distribution of the fluid film is circumferentially averaged to obtain a radial profile as well. Evolutions of the maximum and minimum values of three profiles ($T_{max}$, $T_{s\,max}$, $T_{r\,max}$, $T_{min}$, $T_{s\,min}$ and $T_{r\,min}$) in the TEHD and THD cases are presented in **Fig. 15(b)**. For the curves of the maximum temperatures, it rises from around 60 °C to around 62.5 °C from 0 s to 1.5 s (0 r/min to 1937 r/min), and it keeps going from around 62.5 °C to around 70 °C from 1.5 s to 2.673 s (1937 r/min to 3405 r/min). During the steady-speed period until 8 s, the climbing trend continues. The significant influence of the thermal lag on the heat conduction of the face seal is emphasized according to these phenomena.

Two more features during the steady-speed stage until 8 s can be observed from **Fig. 15(b)**. On the one hand, $T_{max}$ is larger than $T_{s\,max}$ and smaller than $T_{r\,max}$ in both TEHD and THD cases, and the gaps between $T_{r\,max}$ and $T_{s\,max}$ at 8 s are respectively 2.18 °C and 0.45 °C. The temperature difference inside the rotor face is always larger than that inside the stator face especially in the TEHD case, which can be attributed to the difference of the ring structures and the thermal boundaries. On the other hand, the maximum temperature of the TEHD model is larger than that of the THD model, for example, $T_{max}$ at 8 s are 78.35 °C and 73.12 °C. Meanwhile, the minimum temperature of the TEHD model is smaller than that of the THD model. It means that based on the given conditions, lager temperature difference occurs in the TEHD case comparing with the THD case that ignores the solid deformation.

Evolutions of the fluid film temperature distribution from 0 s to 3 s are illustrated in **Fig. 15(c)**. Before 1.5 s, the temperature difference inside the fluid film is really limited due to small temperature increment. After 1.5 s, the variation inside the film becomes more and more evident. The film temperature in the dam area is obviously higher than that in the groove and land areas. The film temperature in the outer-half groove is almost same as the medium temperature, and the temperature becomes larger at the high-pressure area of the grooves. The temperature difference between the inner-half grooves and the lands beside them is not notable. From the results at 2.5 s and 3 s, it can be seen that the temperature level of the TEHD case is larger than that of the THD case, as well as the temperature difference inside the fluid film.

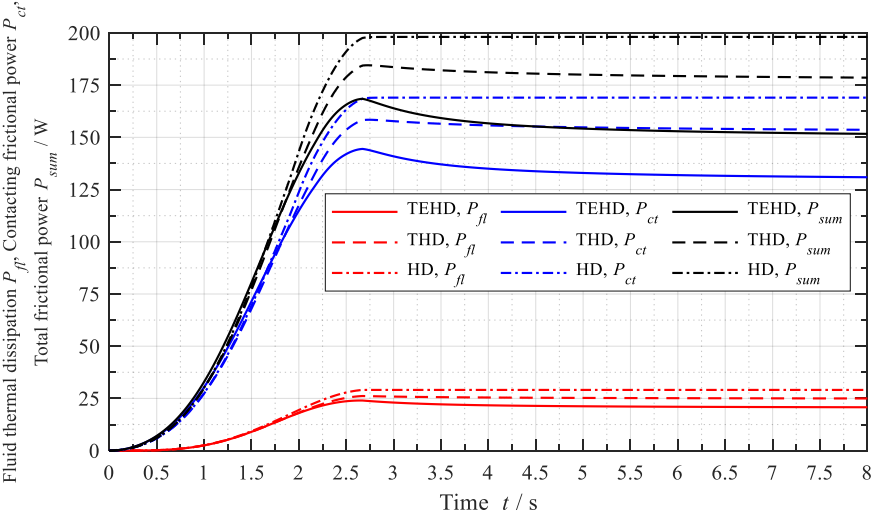

(a)

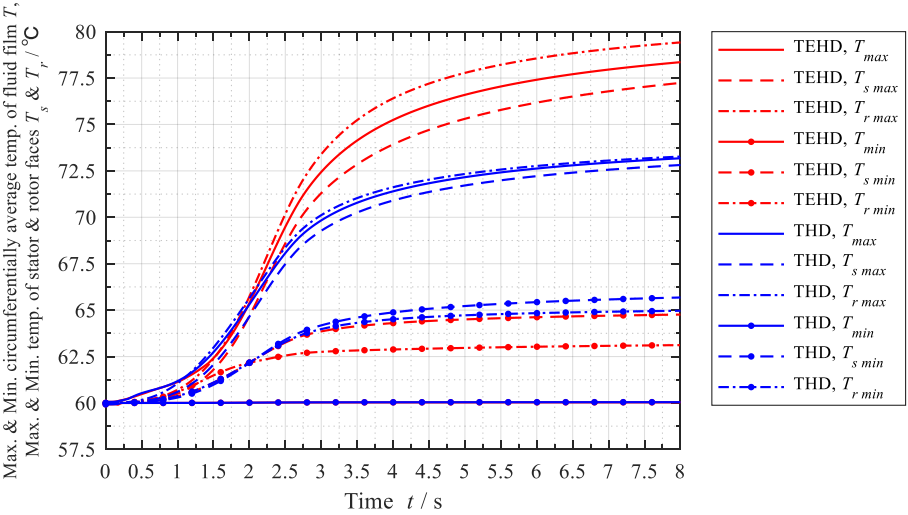

(b)

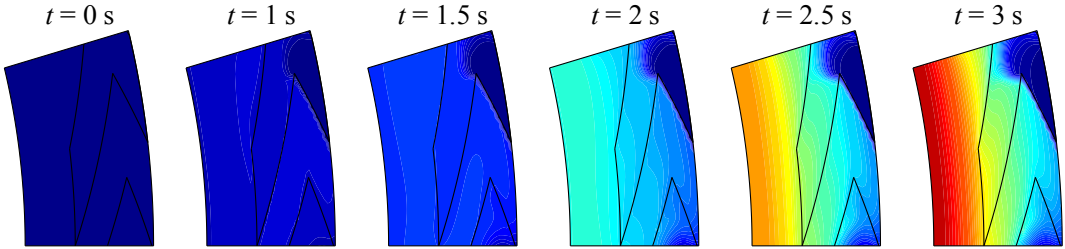

(c-1) the TEHD case

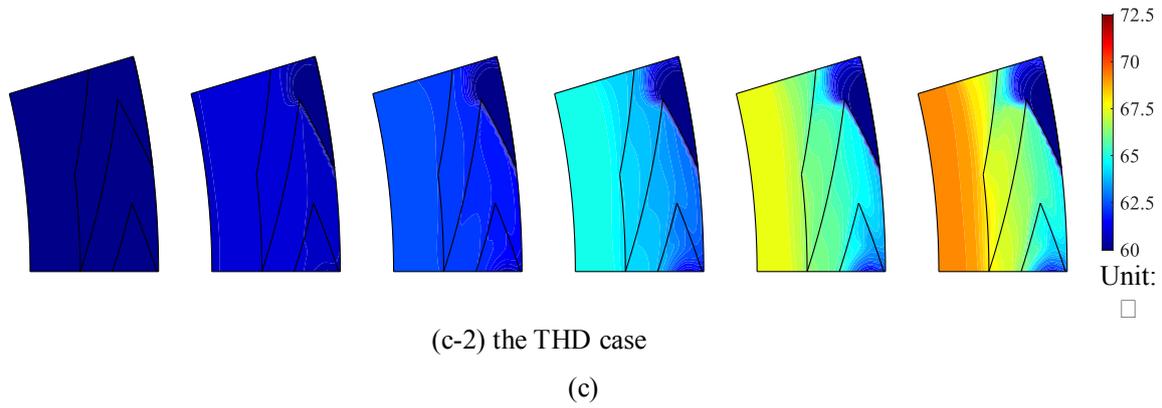

(c-2) the THD case

(c)

**Fig. 15 Thermal properties: (a) evolutions of thermal dissipations of fluid film, contacting frictional powers and total frictional powers, (b) evolutions of maximum and minimum circumferentially average temperatures of fluid film, stator face and rotor face, (c) evolutions of the fluid film temperature distribution (from 0 s to 3 s).**

*5.4.4. Fluid film properties*

The gas volume fraction distributions of the fluid film at 3 s are illustrated in **Fig. 16(a)**, and evolutions of the average gas volume fraction $\alpha_{av}$ are given in **Fig. 16(b)**. Cavitation areas just focus on the very tips of the inner-half grooves and are very limited, because the grooves connect the medium and the pressure diminishing is complemented. The cavitation is negligible for both TEHD and THD cases, but it is more notable in the HD case. As for the average gas volume fraction, differences among three cases start to emerge after 2 s (2787r/min). The peak values are respectively 0.0282, 0.0297 and 0.0367 when reaching the maximum speed. But it smoothly goes down to 0.0250 and 0.0278 in the TEHD and THD cases afterwards. It can be summarized that ignoring the thermal effect can significantly overestimate the cavitation of the liquid film.

The radial mass flow rates at the inner diameter $Q_{m,i}$ are compared in **Fig. 16(c)**. At standstill, $Q_{m,i}$ of the TEHD case is larger than that of the rest, turning out to be 0.566 g/min versus 0.470 g/min. It means when considering the mechanical deformation, larger static leakage happens due to the divergent gap. In three cases, $Q_{m,i}$ declines with speed rising due to the suction effect of the developing low-pressure area in the fluid film. The final leakages of three cases are respectively 0.181 g/min, 0.148 g/min and 0.091 g/min.

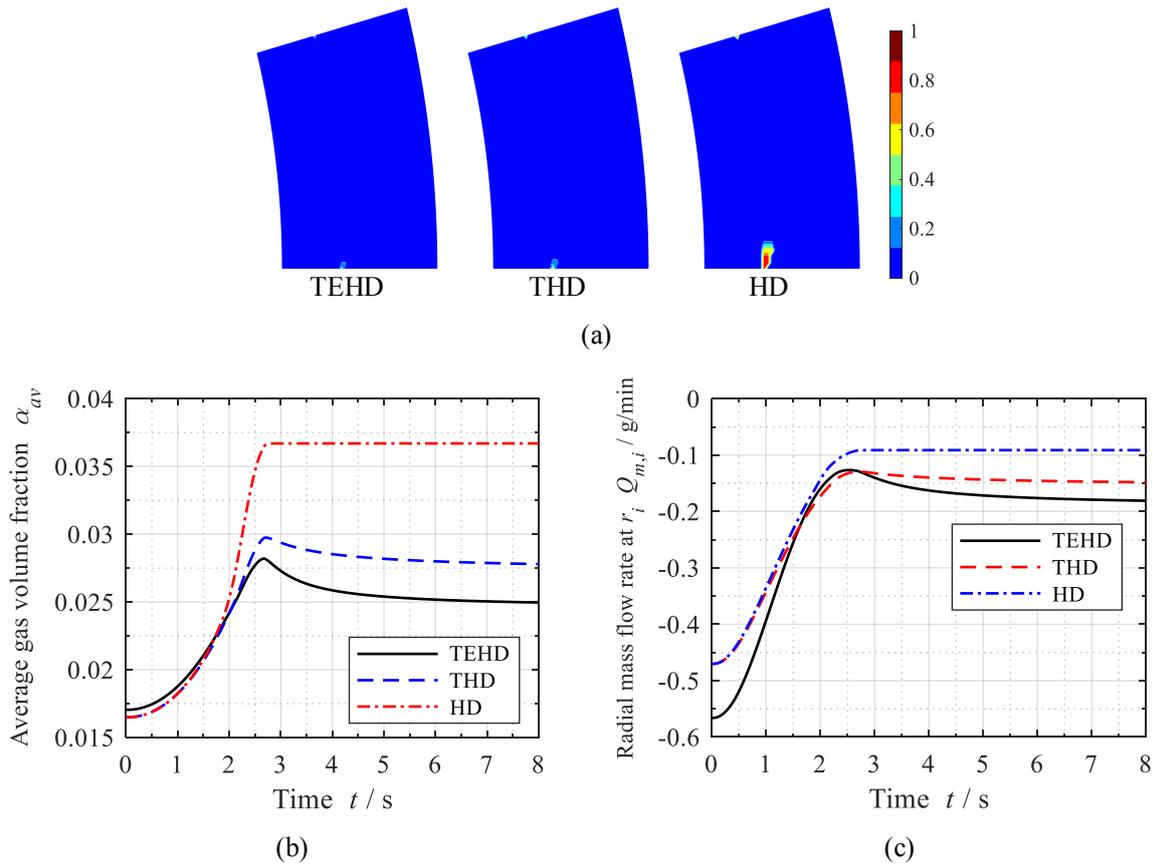

Fig. 16 Fluid film properties: (a) gas volume fraction distributions at $t$ = 3 s, (b) evolutions of average gas volume fraction of the whole fluid film, (c) evolutions of radial mass flow rates at the inner diameter.

## 6. Conclusions

The transient mixed thermo-elasto-hydrodynamic (TEHD) lubrication numerical solution is established for mechanical face seals. Solution of the transient fluid film properties is based on the Finite Volume Method, and the transient solid properties are calculated based on the Duhamel's Principle. An approach of Parallel Dual Time Steps (PDTS approach) is proposed for the explicit time solver of the numerical calculation flow chart. Both of the efficiency and accuracy of the PDTS approach are evaluated by comparing with the reference.

An outer-herringbone-grooved face seal in a start-up stage is studied. The simultaneously existing physical effects of the face expansion and the seal ring movement are successfully simulated with the proposed method. For the given cases, when the viscosity-temperature effect and the convergent-gap forming effect are ignored, the load-carrying capacity of the fluid film and the leakage can be underestimated; the fluid frictional torque, the fluid thermal dissipation and the average gas volume fraction can be overestimated; meanwhile, the maximum contacting pressure which can indicate the possibility of face wear is significantly underestimated. The thermal lag effect works and is necessary to be considered to obtain more realistic evolving process of the seal behavior in the transient or dynamic operations.

It is found that even if the rotation speed reaches the maximum value, the transient thermal performance will induce the corresponding dynamic response. Meanwhile, the contact profile keeps varying. For the giving TEHD case, the maximum contact pressure area moves from the outer diameter to the inner diameter. These findings give explanation to the wear phenomena of seal faces in the field applications.

**CRediT authorship contribution statement**

**Yongfan Li:** Conceptualization, Methodology, Software, Validation, Formal analysis, Investigation, Writing - Original Draft, Writing - Review & Editing, Visualization. **Muming Hao:** Conceptualization, Resources, Writing - Review & Editing, Project administration, Funding acquisition. **Noël Brunetière:** Methodology, Resources, Writing - Review & Editing, Funding acquisition. **Qiang Li:** Methodology, Investigation, Writing - Review & Editing, Visualization, Funding acquisition. **Jiasheng Wang:** Writing - Review & Editing, Visualization; **Baojie Ren:** Writing - Review & Editing, Visualization.

**Declaration of Competing Interest**

The authors declare that they have no known competing financial interests or personal relationships that could have appeared to influence the work reported in this paper.

**Acknowledgements**

This work has been supported by the National Natural Science Foundation of China, and the Grant Numbers are 52176050 and 51975585. This work has been supported by the Project HFZL2023CXY016. This work also pertains to the French government program Investissements Avenir (LABEX INTERACTIFS, reference ANR-11-LABX-0017-01, and EUR INTREE, reference ANR-18-EURE-0010).

**Data availability**

Data will be made available on request.

**Appendix A. FVM Coefficients of the discretized Reynolds equation**

The coefficients in the discretized Reynolds equation (**Eq. (9)**) are given as the following:

$$C_{pN} = \frac{1}{12\Delta r} \cdot \frac{\rho_n C_{pn}}{\mu_n}, \ C_{pS} = \frac{1}{12\Delta r} \cdot \frac{\rho_s C_{ps}}{\mu_s}, \ C_{pE} = \frac{1}{12 r_C \Delta\theta} \cdot \frac{\rho_e C_{pe}}{\mu_e},$$

$$C_{pW} = \frac{1}{12 r_C \Delta\theta} \cdot \frac{\rho_w C_{pw}}{\mu_w}, \ C_{pC} = C_{pN} + C_{pS} + C_{pE} + C_{pW},$$

$$C_{pO} = \rho_C \frac{r_e \omega}{2} \cdot C_{ve} - \rho_W \frac{r_w \omega}{2} \cdot C_{vw} + \rho_C^{k+1} \cdot (\dot{z}_{rC}^{k+1} - \dot{z}_{sC}^{k+1}) \cdot \Delta V^{k+1}$$

in which,

$$C_{pn} = \phi_{pn} h_n^3 A_n, \quad C_{ps} = \phi_{ps} h_s^3 A_s, \quad C_{pe} = \phi_{pe} h_e^3 A_e, \quad C_{pw} = \phi_{pw} h_w^3 A_w,$$

$$C_{ve} = \left(\bar{h}_{Te} + \sigma_e \phi_{se,rs}\right) A_e, \quad C_{vw} = \left(\bar{h}_{Tw} + \sigma_w \phi_{sw,rs}\right) A_w$$

**Appendix B. FVM Coefficients of the discretized energy equation**

The coefficients in the discretized energy equation (**Eq. (11)**) are given as the following:

$$C_{TN}^{k+1} = C_{cpn}^{k+1} S_{-qn}^{k+1}, \quad C_{TS}^{k+1} = C_{cps}^{k+1} S_{+qs}^{k+1},$$

$$C_{TE}^{k+1} = C_{cpe}^{k+1} S_{-qe}^{k+1}, \quad C_{TW}^{k+1} = C_{cpw}^{k+1} S_{+qw}^{k+1},$$

$$C_{TCk} = \frac{c_{pC}^k \rho_C^k \cdot h_C^k \Delta V^{k+1}}{\Delta t^k},$$

$$C_{TO}^{k+1} = \left(H_{ctC}^k - H_{rC}^k - H_{sC}^k\right) \cdot \Delta V^{k+1} + \mu_C^{k+1} (\omega^{k+1})^2 r_C^3 \cdot \Delta\theta \Delta r + \frac{p_C^{k+1} - p_C^k}{\Delta t^k} \cdot h_C^k \Delta V^{k+1}$$

$$+ \left[C_{qn} \cdot (p_n - p_C) + C_{qs} \cdot (p_C - p_s) + C_{qe} \cdot (p_e - p_C) + C_{qw} \cdot (p_C - p_w)\right]^{k+1}.$$

in which,

$$C_{qn} = q_{rn} A_n, \quad C_{qs} = q_{rs} A_s, \quad C_{qe} = q_{\theta e} A_e, \quad C_{qw} = q_{\theta w} A_w,$$

$$C_{\rho n} = \rho_n C_{qn}, \quad C_{\rho s} = \rho_s C_{qs}, \quad C_{\rho e} = \rho_e C_{qe}, \quad C_{\rho w} = \rho_w C_{qw},$$

$$C_{cpn} = c_{pn} C_{\rho n}, \quad C_{cps} = c_{ps} C_{\rho s}, \quad C_{cpe} = c_{pe} C_{\rho e}, \quad C_{cpw} = c_{pw} C_{\rho w},$$

$$S_{+qn} = \frac{1 + sign(q_{rn})}{2}, \quad S_{-qn} = \frac{1 - sign(q_{rn})}{2}, \quad S_{+qs} = \frac{1 + sign(q_{rs})}{2}, \quad S_{-qs} = \frac{1 - sign(q_{rs})}{2},$$

$$S_{+qe} = \frac{1 + sign(q_{\theta e})}{2}, \quad S_{-qe} = \frac{1 - sign(q_{\theta e})}{2}, \quad S_{+qw} = \frac{1 + sign(q_{\theta w})}{2}, \quad S_{-qw} = \frac{1 - sign(q_{\theta w})}{2}.$$